\documentclass{article}
\usepackage{bm}
\usepackage[numbers,sort&compress]{natbib}
\usepackage{multibib}
\usepackage{amsmath}
\usepackage{amssymb}
\usepackage{amsfonts}
\usepackage{graphicx}
\usepackage{textcomp}
\newcites{S}{Supporting Information References}

\title{Satellite observations of nitrous oxide emissions from adipic acid production}
\author{
    Nicholas Balasus* and Daniel H.~Cusworth\\[1ex]
    \footnotesize *Corresponding author: \texttt{nicholas@carbonmapper.org}\\[1ex]
    \footnotesize Carbon Mapper, Pasadena, CA, USA
}
\date{\today}

\begin{document}

\maketitle

\begin{abstract}
The production of adipic acid generates nitrous oxide as a by-product, which, if emitted, very effectively warms the climate and depletes the stratospheric ozone layer. There exists cost-effective technology to mitigate these nitrous oxide emissions from adipic acid production, but the degree of implementation of this technology, most notably at sites in China, is uncertain. Using satellite observations from summer 2026, we quantify nitrous oxide emissions from three adipic acid production sites in China, finding emissions that total over half a million tonnes per year, a rate that implies limited emissions abatement. With respect to climate change, eliminating the annual emissions from just these three sites would be equivalent to eliminating the annual emissions from 49 coal-fired power plants or 43 million gas-powered vehicles in the United States.
\end{abstract}

\section{Introduction}
Nitrous oxide (N$_2$O) emissions warm the climate 273 times more effectively than carbon dioxide on a per-mass basis \cite{forster2021}, and they contribute significantly to stratospheric ozone depletion \cite{ravishankara2009}. The production of adipic acid, which primarily serves as a feedstock for nylon 6,6 production, has long been recognized as a globally significant source of N$_2$O \cite{thiemens1991}, and one that continues to grow in recent years, especially in China \cite{davidson2023}. The Intergovernmental Panel on Climate Change recommends an emission factor of 0.3 kg N$_2$O per kg adipic acid produced \cite{ipcc2006}, which follows from the ratio of their molar masses, as they are produced in approximately stoichiometric quantities \cite{thiemens1991}. Although agriculture dominates global anthropogenic N$_2$O emissions \cite{tian2024}, agricultural emissions are intrinsically more challenging to mitigate compared to industrial sources like adipic acid \cite{aziz2024}.

There are estimated to be fewer than 50 adipic acid production sites worldwide \cite{davidson2023,hasanbeigi2023}. Starting in the 1990s, many sites in the United States and Europe, where production was initially concentrated, implemented thermal destruction and catalytic decomposition technology, allowing them to destroy more than 90\% of the N$_2$O produced \cite{reimer2000}. This technology remains cost-effective today \cite{davidson2023}.

China is now the global leader in adipic acid production \cite{hasanbeigi2023}. The degree to which adipic acid production sites in China are abating N$_2$O emissions, however, remains uncertain \cite{mckenna2020}. Select sites in China previously abated a verified near 100\% of their N$_2$O emissions, motivated by financial incentives from the Clean Development Mechanism (CDM) from the United Nations, a carbon credit program \cite{schneider2010}. However, the design of the CDM was such that it distorted the global adipic acid market; selling the carbon credits through the program became more lucrative than selling adipic acid, shifting production toward CDM sites and thus likely resulting in a net increase in global greenhouse gas emissions \cite{schneider2010}. In 2011, the European Union approved a ban on the use of credits from CDM adipic acid projects in the European Union Emissions Trading System \cite{eu2011}, and it soon became no longer profitable for the CDM adipic acid sites to abate their N$_2$O emissions \cite{mckenna2020}. Since then, it is unclear if enough incentives, financial or otherwise, exist in China to motivate emissions abatement \cite{mckenna2020}.

To address this uncertainty, we use satellite observations to locate and quantify N$_2$O emissions from adipic acid production facilities in China. Satellite observations offer an avenue to map facility-scale pollution emissions, as demonstrated for example by Tanager-1 and EMIT, which have been used for mapping methane, carbon dioxide, carbon monoxide, and ammonia emissions using gas absorption features in the shortwave infrared \cite{duren2025,thorpe2023,ruzicka2026,balasus2026}. This study expands these shortwave infrared imaging spectroscopy capabilities to now include N$_2$O.

\section{Materials and Methods}
We use radiance spectra from 2050 to 2350 nm from the Tanager-1 pushbroom imaging spectrometer, which offers 5 nm spectral sampling and 30 m spatial resolution \cite{duren2025}, to calculate N$_2$O vertical column enhancements using a matched filter retrieval with scene-specific unit absorption spectra (Section~\ref{sec:S1}). We then use the integrated mass enhancement method to infer emission rates (Section~\ref{sec:S2}). This combination of methods has performed well with Tanager-1 data in blinded methane controlled release studies \cite{reuland2026}, which we anticipate to be readily adaptable to N$_2$O given that it is also effectively inert at these scales. Similar methods also perform well for the quantification of carbon dioxide emissions, as validated by continuous emissions monitoring systems \cite{nelson2024,kim2025}.

We also use data from the EMIT imaging spectrometer onboard the International Space Station \cite{thorpe2023}, which offers more days of observations for our sites of interest than Tanager-1, but also with lower sensitivity to N$_2$O, owing to its coarser spectral sampling (7 nm) and spatial resolution (60 m). The noise in the N$_2$O retrievals from EMIT challenges our quantification algorithms, particularly in the delineation of the plume for sites with lower emission rates, so we opt to wind-rotate and time-average the data from EMIT (Section~\ref{sec:S3}).

\begin{figure}
    \centering
    \makebox[\textwidth][c]{\includegraphics[width=1.3\textwidth]{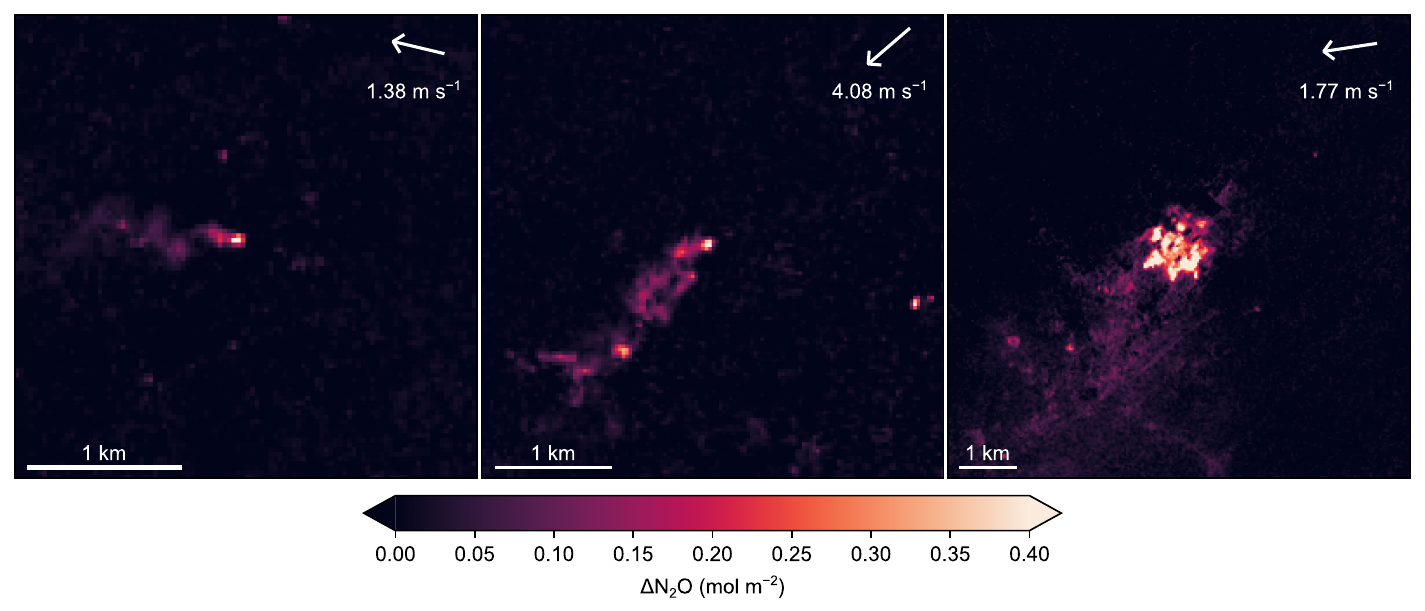}}%
    \caption{N$_2$O vertical column enhancements retrieved using Tanager-1 data. From left to right, the sites shown are Taiyuan, China on 26 July 2026 (37.574\textdegree N, 112.284\textdegree E); Dezhou, China on 9 August 2026 (37.470\textdegree N, 116.227\textdegree E); and Chongqing, China on 1 September 2026 (29.601\textdegree N, 107.555\textdegree E). The 10-meter wind speed and direction from ERA5 are inset.}
    \label{fig:1}
\end{figure}

\section{Results and Discussion}
Figure~\ref{fig:1} shows Tanager-1 observations of N$_2$O over adipic acid production sites in Taiyuan, Dezhou, and Chongqing in China. These sites were identified using reported facility names and approximate locations \cite{mckenna2020}. From these observations made in summer 2026, we estimate emissions of 2.15 (1.06--3.22) t h$^{-1}$ from Taiyuan, 19.6 (17.7--21.2) t h$^{-1}$ from Dezhou, and 55.6 (35.9--75.0) t h$^{-1}$ from Chongqing, for an annualized sum of 0.68 million tonnes per year. Especially for Dezhou and Chongqing, these large emission rates imply that emissions abatement is limited considering their expected adipic acid production rates.

We check the representativeness of our single-pass Tanager-1 observations using observations from EMIT that are cloud-free over our site of interest and without major geolocation issue. In the case of Taiyuan and Dezhou, we use scenes from the years 2025 and 2026, such that we can expect adipic acid production to be similar to our Tanager-1 observation time, and wind-rotate and then time-average the retrieved N$_2$O column enhancements from those days (Figure~\ref{fig:S1}). For Taiyuan, the composite of nine observation days (covering six unique calendar months) gives an emission rate of 1.93 t h$^{-1}$, while for Dezhou, the composite of four observation days (covering three unique calendar months) gives an emission rate of 19.2 t h$^{-1}$, both consistent with our Tanager-1 observations. For Chongqing, there is only one quality observation from EMIT, which is on 28 April 2024. Using the same single-pass quantification techniques we use for Tanager-1 (Section~\ref{sec:S2}), we quantify emissions of 59.9 (54.7--66.3) t h$^{-1}$ on this day (Figure~\ref{fig:S2}), again consistent with our Tanager-1 observation.

This set of three sites is unlikely to be a comprehensive set of adipic acid production sites that can be observed by the class of sensors that includes EMIT and the growing Tanager constellation. For instance, Tanager-1 observations on 7 July 2025 and 9 April 2026 identified two N$_2$O emission sources in Pingdingshan, China (Figure~\ref{fig:S3}), a site that participated in the CDM. However, a subsequent 28 June 2026 observation only detected a plume at the southern site, and another on 6 September 2026 only detected a plume at the northern site. This variability in detections could represent changing production or implementation of emission abatement technology, and because of this uncertainty, we opt not to include these locations in our reported emissions total. There are at least six other adipic acid production sites in China past those explored here \cite{mckenna2020}, though with uncertain exact production site locations and abatement implementation.

Recent bottom-up estimates of global N$_2$O emissions from adipic acid production \cite{davidson2023} are less than half our total of 0.68 million tonnes per year from these three sites, likely due to a combination of outdated production data and uncertainty surrounding abatement. As such, the role of adipic acid in global N$_2$O budgets, and thus industrial sources more broadly, is likely underestimated.

Using a global warming potential of 273 for N$_2$O \cite{forster2021}, and equivalences of 3.79 million tonnes CO$_2$-eq per coal-fired power plant per year and 4.29 tonnes CO$_2$-eq per gas-powered vehicle per year from the U.S. Environmental Protection Agency \cite{epaghg}, our sum of emissions from the three sites in Taiyuan, Dezhou, and Chongqing can be equated to emissions from approximately 49 coal-fired power plants or 43 million gas-powered vehicles annually. Given the existing technology that can be used to abate these emissions, and especially when considering that adipic acid production continues to rise in China \cite{liang2024}, we view abating these emissions as an urgent target in the context of mitigating climate change.

\section*{Acknowledgments}
We acknowledge the generous support of the philanthropic donors to Carbon Mapper, including the High Tide Foundation, Grantham Foundation for the Protection of the Environment, Bloomberg Philanthropies, and other contributors. We thank Joseph Heckler for his help in collecting the Tanager-1 data.

\clearpage

\bibliographystyle{unsrtnat}
\bibliography{references}

\clearpage
\setcounter{page}{1}
\setcounter{figure}{0}
\setcounter{equation}{0}
\setcounter{subsection}{0}
\renewcommand\thepage{S\arabic{page}}
\renewcommand{\thefigure}{S\arabic{figure}}
\renewcommand{\theequation}{S\arabic{equation}}
\renewcommand{\thesubsection}{S\arabic{subsection}}

\section*{Supporting Information}

\subsection{Vertical Column Enhancement Retrieval}\label{sec:S1}
\begin{figure}
    \centering
    \includegraphics[width=\textwidth]{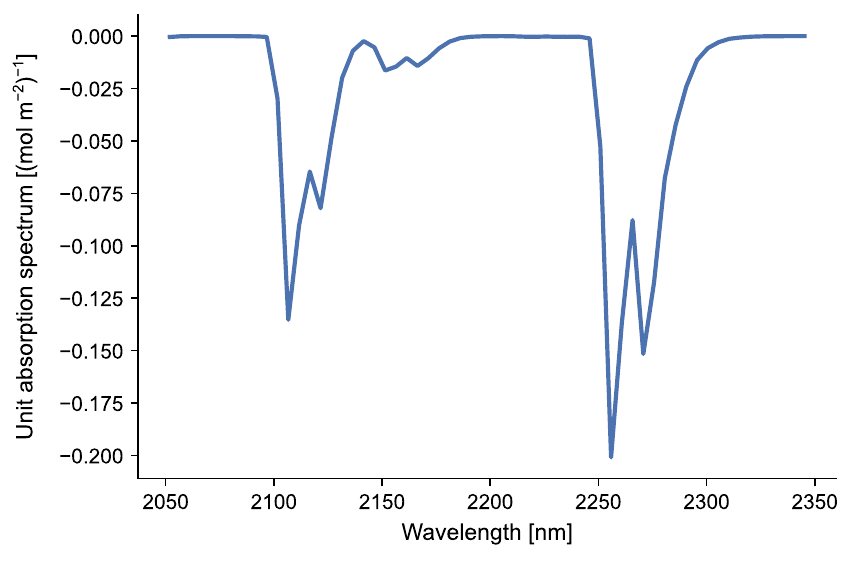} 
    \caption{N$_2$O unit absorption spectrum example for Tanager-1. Unit absorption spectra are generated per scene, and this example corresponds to the observation for the facility in Dezhou on 9 August 2026.}
    \label{fig:S4}
\end{figure}

Tanager-1 is a pushbroom imaging spectrometer with a spectral range of 400 to 2500 nm with 5 nm spectral sampling and 30 m nadir spatial resolution. Using calibrated radiance spectra $\bm{x}\in\mathbb{R}^n$ from Tanager-1 (using the $n$ channels within 2050 to 2350 nm), we estimate N$_2$O vertical column concentration enhancements $\hat{\alpha}$ (mol m$^{-2}$) using the matched filter algorithm, which follows from a least-squares fit using a first-order Taylor expansion of the Beer-Lambert law \citeS{scheidweiler2026}.
\begin{equation}
    \hat{\alpha} = \frac{(\bm{x}-\bm{\mu})^T\bm{\Sigma}^{-1}\bm{t}}{\bm{t}^T\bm{\Sigma}^{-1}\bm{t}}
\end{equation}
We implement a column-wise matched filter such that $\bm{\mu}$ is the mean of all spectra from the same detector in a scene and the covariance $\bm{\Sigma}\in\mathbb{R}^{n\times n}$ is calculated from that same sample. Based on an initial pass of the matched filter, for these background models of $\bm{\mu}$ and $\bm{\Sigma}$, we exclude spectra that correspond to an $\hat{\alpha}$ that is four standard deviations above the mean in the scene then rerun the retrieval. The target signature $\bm{t}$ is the product of $\bm{\mu}$ and a unit absorption spectrum $\bm{s}$, the latter describing the expected change in at-sensor radiance for a given N$_2$O enhancement. We calculate $\bm{s}$ for each scene using a non-scattering radiative transfer model of the at-sensor radiance $\bm{I}$ at 0.001 nm resolution.
\begin{equation}
    \bm{I} = \frac{\bm{E}_0\cdot \bm{\rho}\cdot \cos\theta_s}{\pi}\cdot \exp\left(-a\cdot\sum_g\sum_lu_{g,l}\cdot\bm{k}_{g,l}\right)
\end{equation}
Here, $\bm{E}_0$ is the solar irradiance \citeS{coddington2023}, $\bm{\rho}$ is the surface reflectance (assumed 1.0), $\theta_s$ is the solar zenith angle, $a$ is the geometric air mass factor ($\sec\theta_s + \sec\theta_v$) using the viewing zenith angle $\theta_v$, and $u$ and $\bm{k}$ are the vertical column concentrations and absorption coefficients for each gas $g$ and vertical layer $l$. We query ERA5-Land for the surface pressure at the scene-average time and location, which we use to define the 30-layer pressure grid of our model, using 31 equidistant pressure levels from the surface up to 1 Pa. We interpolate temperature and water vapor mixing ratio profiles from ERA5 to this pressure grid, and do the same for mixing ratios of N$_2$O, CH$_4$, CO, NH$_3$, and CO$_2$ from the U.S. standard atmosphere (with scaling based on modern surface concentrations of 0.33, 1.9, and 420 ppm for N$_2$O, CH$_4$, and CO$_2$ respectively) \citeS{anderson1986}. Absorption coefficients are from HITRAN2024 \citeS{gordon2026} using Voigt line shapes at the temperature and pressure of each layer. We run the radiative transfer model twice for each scene, once with background conditions as described above, and then again adding an $\alpha = 0.01$ mol m$^{-2}$ enhancement of N$_2$O in the surface layer, giving spectra $\bm{I}_{\mathrm{bkg}}$ and $\bm{I}_{\mathrm{enh}}$. We convolve these with the instrument spectral response function, assumed Gaussian for each channel with calibrated band widths, to yield $\bm{L}_{\mathrm{bkg}}$ and $\bm{L}_{\mathrm{enh}}$, which we use to generate the unit absorption spectra $\bm{s}$.
\begin{equation}
    \bm{s} = \frac{\ln(\bm{L}_{\mathrm{enh}})-\ln(\bm{L}_{\mathrm{bkg}})}{\alpha}
\end{equation}
An example N$_2$O unit absorption spectrum for Tanager-1 is shown in Figure~\ref{fig:S4}.

\subsection{Single-Pass Observations Emission Rate Calculation}\label{sec:S2}
We use the integrated mass enhancement method \citeS{frankenberg2016,varon2018} to calculate an emission rate $Q$ (t h$^{-1}$) from the vertical column enhancements $\hat{\alpha}$ (mol m$^{-2}$) for a given plume. We initialize a binary plume mask for pixels within $\pm$1 km of the source using enhancement thresholding of four times the scene-wide $\hat{\alpha}$ standard deviation. We refine this mask with a 5-pixel window median filter and a 1-pixel dilation and use the largest remaining contiguous area as the final plume mask.
\begin{equation}
    \label{eqn:S4}
    Q = \frac{U}{L}\cdot\sum_i \hat{\alpha}_i\cdot A_i\cdot M
\end{equation}
Here, $U$ is the 10-meter wind speed (m h$^{-1}$), $L$ is the plume length (m) defined as the largest dimension of the plume mask convex hull, $A_i$ is the area of each pixel $i$ contained within the plume mask (m$^2$), and $M$ is the N$_2$O molar mass (t mol$^{-1}$). We source the 10-m wind speed from ERA5. For our best estimate of the emission rate and its uncertainty, we use a Monte Carlo approach. For 1000 draws, we add pixel-wise Gaussian noise to our scene based on the scene-wide $\hat{\alpha}$ standard deviation, repeat our masking procedure, then calculate the emission using a wind speed sampled from a Gaussian distribution with the mean from ERA5 and the standard deviation from the ERA5-Ensemble spread. From the resulting distribution, we report the median as our best estimate alongside a 95\% confidence interval from the 0.025 to 0.975 quantiles.

\subsection{EMIT Wind Rotation}\label{sec:S3}
EMIT covers the same spectral range as Tanager-1, but with coarser spectral sampling (7 nm) and spatial resolution (60 m). To create the time-averaged plumes from EMIT, we collect scenes that we judge to be cloud-free over the site of interest and without major geolocation errors. We then retrieve the column enhancements as in Section~\ref{sec:S1}. We apply the EMIT geometric lookup table to map our retrievals to a regular latitude and longitude grid, then project the corners of each pixel to an azimuthal equidistant projection centered on the site. Based on the ERA5 wind direction at the site at the observation time, we rotate the pixel corner coordinates $x$ and $y$ by $\theta$ to align with westerly winds.
\begin{equation}
    x'=x \cdot \cos \theta - y \cdot \sin \theta
\end{equation}
\begin{equation}
    y'=x \cdot \sin \theta + y \cdot \cos \theta
\end{equation}
Following this wind rotation, we construct polygons from the rotated pixel corner coordinates and average these polygons into 50 m $\times$ 50 m grid cells within $\pm$2 km of the site, using a weighted average based on the intersecting area of the observation polygons and the target grid cell. We then initialize a plume mask using pixels above the 95$^{\text{th}}$ percentile of values in the resulting 4 km $\times$ 4 km image, less restrictive than the single-pass case due to a smaller sample size, then apply a 5-pixel window median filter and a 1-pixel dilation. We use this mask to calculate the emission using the integrated mass enhancement method.

\begin{figure}
    \centering
    \includegraphics[width=\textwidth]{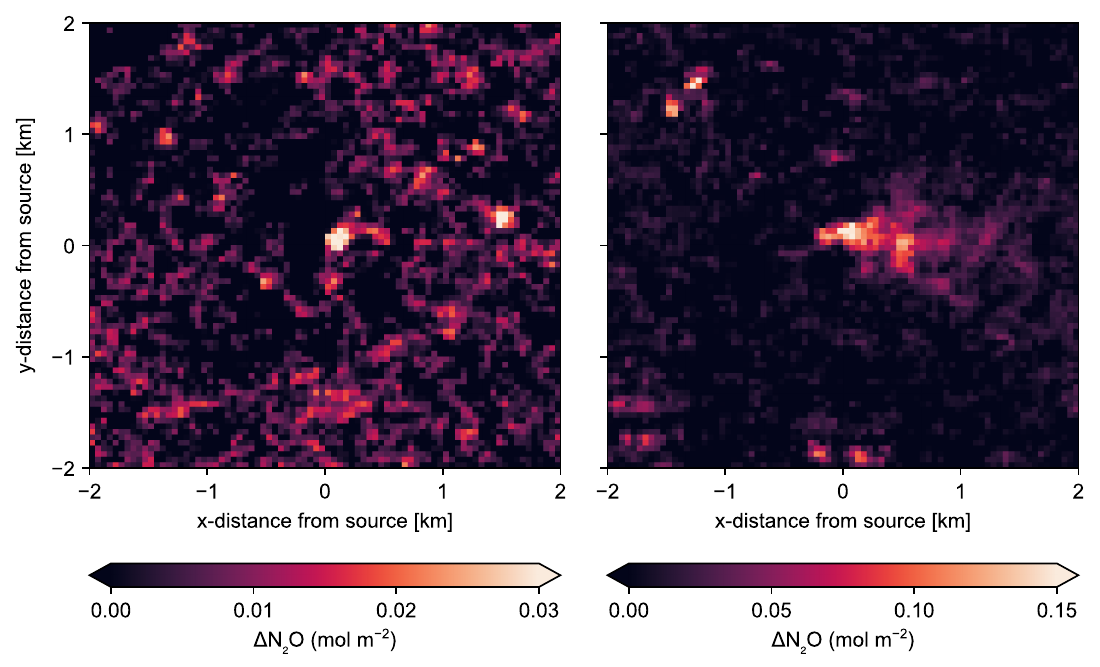} 
    \caption{Time-averaged N$_2$O vertical column enhancements retrieved using EMIT data in Taiyuan, China (left) and Dezhou, China (right). The site is at the origin of the images, which is (37.574\textdegree N, 112.284\textdegree E) for Taiyuan and (37.470\textdegree N, 116.227\textdegree E) for Dezhou, and observations are rotated to simulate a westerly wind based on the ERA5 wind direction at observation time. The Taiyuan image contains the average of nine scenes (30 March 2025, 3 April 2025, 4 June 2025, 10 August 2025, 25 September 2025, 30 November 2025, 29 March 2026, 11 June 2026, and 15 June 2026) and the Dezhou image contains four scenes (15 June 2025, 18 April 2026, 22 April 2026, and 28 May 2026).}
    \label{fig:S1}
\end{figure}

\begin{figure}
    \centering
    \includegraphics[width=\textwidth]{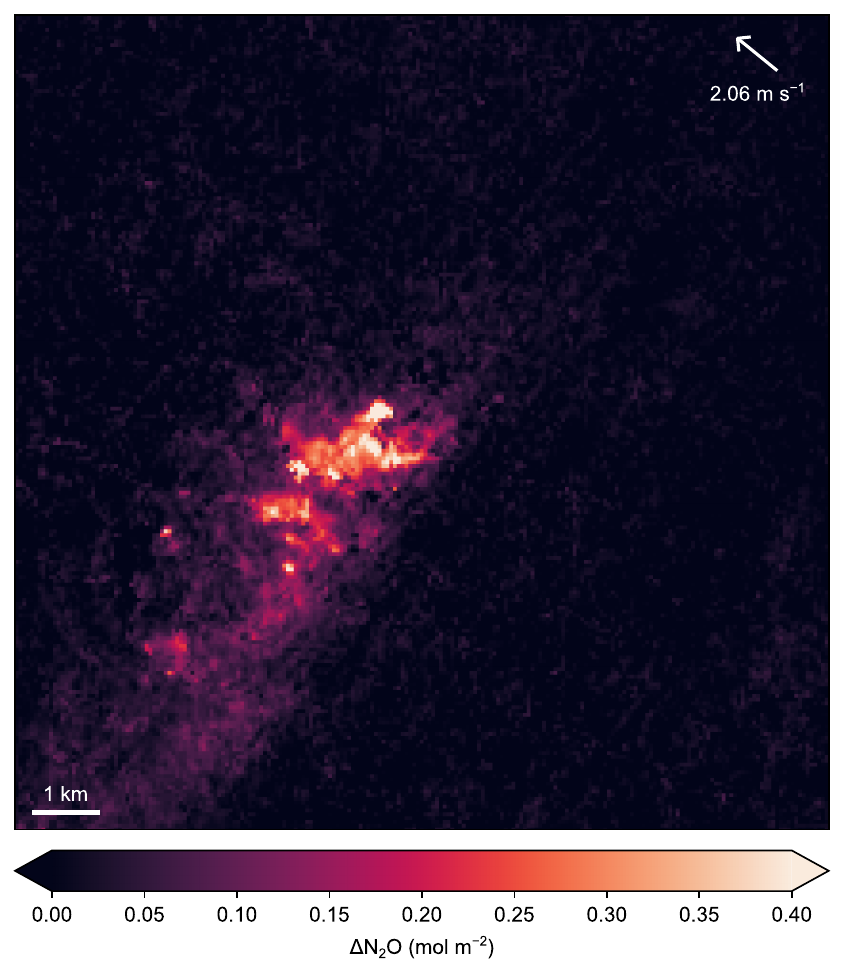} 
    \caption{N$_2$O vertical column enhancements retrieved using EMIT data in Chongqing, China on 28 April 2024. The image is centered on the coordinates (29.601\textdegree N, 107.555\textdegree E). The ERA5 10-meter wind speed and direction is inset.}
    \label{fig:S2}
\end{figure}

\begin{figure}
    \centering
    \includegraphics[width=\textwidth]{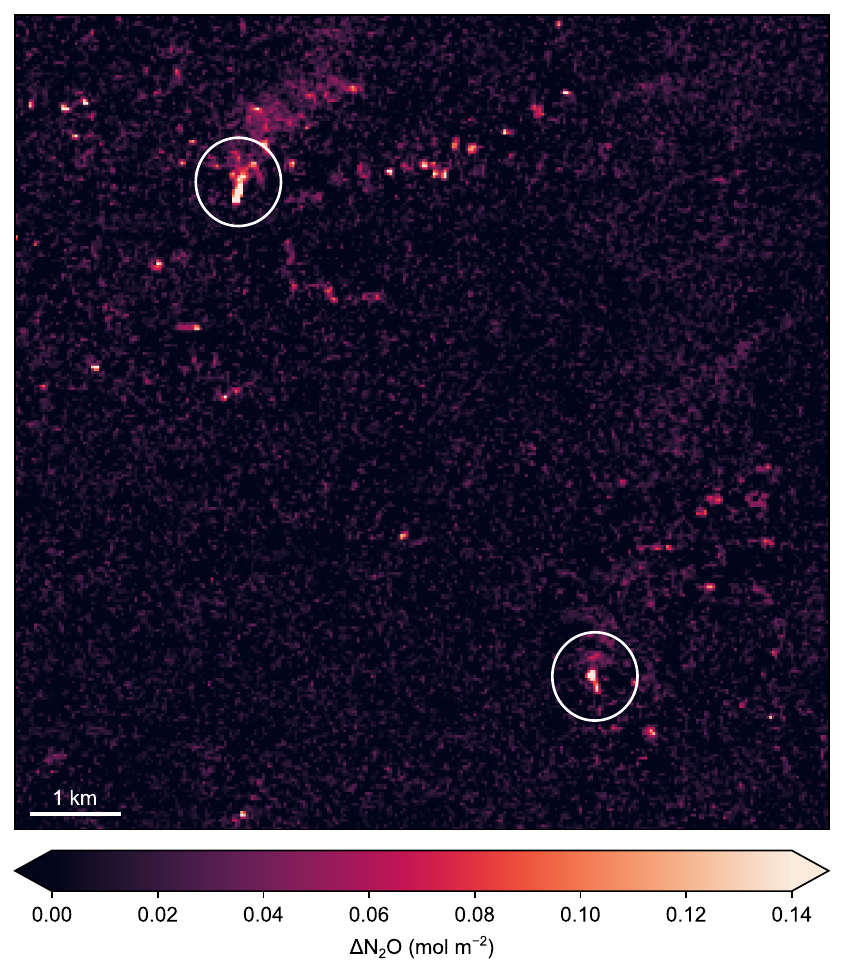} 
    \caption{N$_2$O vertical column enhancements retrieved using Tanager-1 data in Pingdingshan, China on 7 July 2025. The two identified emission sources are circled in white. The image is centered on the coordinates (33.706\textdegree N, 113.404\textdegree E).}
    \label{fig:S3}
\end{figure}

\clearpage
\bibliographystyleS{unsrtnat}
\bibliographyS{references}

\end{document}